\PassOptionsToPackage{dvipsnames}{xcolor}
\documentclass[conference]{IEEEtran}

\usepackage{amsmath,amssymb}
\usepackage{cite}
\usepackage{textcomp}
\usepackage{graphicx}
\graphicspath{{figures/}}
\usepackage{tabularx}
\usepackage{float}      %
\usepackage[section]{placeins} %
\usepackage{booktabs} %
\usepackage{array}   %
\usepackage[linesnumbered,ruled,vlined]{algorithm2e}
\usepackage{listings}
\usepackage{tikz}
\usepackage{comment}
\usepackage[caption=false,font=footnotesize]{subfig}
\usepackage{xurl}
\usepackage{microtype}

\usepackage{xcolor}
\colorlet{TSseekMotivationProposal}{black}
\colorlet{TSseekRevisionRoundTwo}{black}
\definecolor{TSseekCurrentRevision}{HTML}{1B6E3C}
\newsavebox{\TSseekWholeFormulaBox}
\definecolor{TSseekQueryWildcard}{HTML}{9955FF}
\newcommand{\TSseekWholeRunningQuery}{%
  \color{black}%
  Q_{\mathrm{whole}}=\big\langle
  \underbrace{\textcolor{blue}{[100,200]}}_{\text{(a)}},\;
  \underbrace{\big|\textcolor{blue}{[200,800]}\,\frac{\textcolor{red}{+}}{\textcolor{red}{[100,200]}}\,(6)\big|}_{\text{(b)}},\;
  \underbrace{\big|\textcolor{blue}{[200,800]}\,\overset{\textcolor{red}{-}}{\text{---}}\,([5,7])\big|}_{\text{(c)}},\;
  \underbrace{\textcolor{red}{W}}_{\text{(d)}}
  \big\rangle .%
}

\usepackage[normalem]{ulem}   %

\newif\ifshowEARcomments
\showEARcommentstrue

\newif\ifshowEltabakhcomments
\showEltabakhcommentsfalse
\newcommand{\eltabakhcomment}[1]{\ifshowEltabakhcomments#1\fi}

\newif\ifshowAdditionalInfo
\showAdditionalInfofalse
\newcommand{\additionalinfo}[1]{\ifshowAdditionalInfo\par\begingroup\noindent\textbf{Additional information.}\par#1\par\endgroup\fi}

\usepackage[hidelinks]{hyperref}
\title{\textsc{TSseek}: Pattern-Based Search for Distributed Time Series Datasets}

\author{
\IEEEauthorblockN{Xiaoshuai Li$^+$, Khalid Alnuaim$^+$, Mohamed Y. Eltabakh$^*$, and Elke A. Rundensteiner$^+$}
\IEEEauthorblockA{$^+$Worcester Polytechnic Institute, Worcester, MA, USA\\
$^*$Qatar Computing Research Institute, Doha, Qatar\\
\{xli3,~kalnuaim,~rundenst\}@wpi.edu \quad meltabakh@hbku.edu.qa}
}

\begin{document}

\maketitle

\begin{abstract}
In applications such as household energy monitoring and ECG analysis, analysts may search for time series that satisfy constraints on trends, value ranges, and durations.
To support such queries at scale, we propose \textsc{TSseek}, a distributed processing engine for time series pattern queries.
Its regex-inspired language enables users to compose trends, value ranges, and wildcards with constraints on changes between readings and the duration, position, order, and spacing of patterns.
During indexing, \textsc{TSseek} approximates each stored time series with error-bounded line segments, records their slope and value-range metadata, and builds \textsc{TSseek-X}, a distributed spatial index over the segments.
At query time, \textsc{TSseek} rewrites the query and uses selectivity estimates to plan the index lookup. During execution, index lookup uses spatial probes derived from query constraints over reading positions and values to select candidate series from \textsc{TSseek-X} while guaranteeing to retain every true match. Finally, exact refinement checks each candidate against the full query using its original readings.
Experiments on four datasets of up to 200 million time series (25.6 billion readings) show  $\boldsymbol{20\times}$ speedup  over a distributed exact full scan using the same matcher.
{In the whole-matching comparison, the adapted baselines each reach at most 57\% recall, while  \textsc{TSseek}  returns the exact answer set.}
\end{abstract}


\colorlet{TSseekCurrentRevision}{black}

\section{Introduction}
\label{sec:introduction}

\begingroup\color{TSseekRevisionRoundTwo}
\subsection{Background and Motivation}

Time series search methods typically rely on distance-based retrieval~\cite{ding2008querying}. Given an example time series as the query, these methods retrieve similar sequences from a database under a specified distance measure~\cite{KeoghRatanamahatana2005ExactDTW,10.1007/s10618-024-01005-2,zhang2019tardis,wang2021seanet,ShiehKeogh2009}.
However, analysts often seek patterns that cannot be naturally expressed as a single example sequence. Instead, they may wish to specify constraints on a pattern's structure, such as its trends, value ranges, or the fixed and variable durations of its constituent phases. These constraints characterize the properties that every valid match must satisfy, rather than defining a particular numerical sequence against which candidates are compared. Representing such requirements through example sequences is both cumbersome and fundamentally inadequate. Even a simple condition on two consecutive readings, such as $x_1 < x_2$, corresponds to infinitely many valid value combinations, making explicit enumeration impossible. Consequently, users need an expressive pattern-based query language that enables
them to specify
structural and semantic constraints directly, without having to encode them as example sequences.

Existing systems cover only parts of this design space. \textcolor{TSseekCurrentRevision}{Distance-based methods use measures such as DTW~\cite{berndt1994dtw,KeoghRatanamahatana2005ExactDTW} and Euclidean distance~\cite{ding2008querying}. They can summarize time series using compact representations such as PAA~\cite{keogh2001dimensionality} and SAX~\cite{lin2007experiencing}.} Systems such as \textsc{TARDIS}~\cite{zhang2019tardis} and \textsc{SEAnet}~\cite{wang2021seanet} then accelerate this similarity search through distributed indexing and learned representations, respectively.
\begingroup\color{TSseekCurrentRevision}
\textsc{T-ReX}~\cite{T_Rex} supports explicit pattern conditions and optimizes their evaluation within each time series at query time, without a persistent dataset-level index. 
The following example motivates why an analyst may specify a pattern-based query on time series datasets.

\par\endgroup
\endgroup

{
\textbf{Motivating Example.}
{\itshape\color{TSseekCurrentRevision}
A household notices brief bursts on its electricity monitor when demand is otherwise low and asks its electricity supplier for help identifying what might be drawing power. The supplier searches its archive of whole-house power readings for records that begin at low demand, rise, and then decline. The supplier sets a fixed rise duration and bounds each increase to distinguish a buildup from a sudden jump. Both phases must stay within the power range of the reported bursts. The query allows several decline lengths and leaves subsequent readings unconstrained. Example~1 shows how these conditions are expressed with numerical bounds.

The supplier runs the same query across monitored homes to find more periods to investigate and checks which devices were active where appliance-level readings are available. At REFIT's eight-second sampling interval~\cite{murray2017refit}, two years of continuous readings from 3,000 homes would give the supplier about 185 million non-overlapping 128-reading series to search. Finding every qualifying record across an archive of this size makes it important to limit how much raw data each query reads.
\par}
}

\subsection{Challenges}
\begingroup
\textcolor{TSseekCurrentRevision}{Answering such pattern queries exactly at this scale raises two challenges.}

\subsubsection{Compatible Feature Extraction and Indexing}
With a numerical query sequence, summaries such as PAA~\cite{keogh2001dimensionality} and SAX~\cite{lin2007experiencing} can be computed for both the query and the stored series. A complex pattern query, as motivated above, provides conditions on values, trends, and durations rather than a mere numerical sequence to summarize.
The challenge stems from the need to reconcile two fundamentally distinct object types, time-series data and abstract pattern conditions, within a single shared representation space. 
This requires preserving the semantics of both while enabling efficient indexing and  selective retrieval based on their alignment in that common space.

\subsubsection{Exact Query Processing at Scale}
While pattern-search engines such as \textsc{T-ReX}~\cite{T_Rex} optimize evaluation of a pattern within a single time series, \textcolor{TSseekCurrentRevision}{large-scale retrieval requires deciding which series to read and check against the full query.}  Pattern queries often admit multiple valid temporal alignments due to flexible phase durations and transition boundaries, making it difficult to rule out candidate series without accessing raw data. The challenge is therefore to find effective processing strategies that aggressively prune the search space and reduce communication across nodes while guaranteeing exact retrieval without exhaustive scanning.
\par\endgroup

\color{black}

\subsection{\textcolor{TSseekRevisionRoundTwo}{Proposed Approach}}

\textcolor{TSseekCurrentRevision}{We introduce \textbf{\textsc{TSseek}}, a framework for exact pattern search over large distributed time series datasets. Users specify queries in a regex-inspired language that combines numerical values, ranges, trends, and wildcards with duration and step-size constraints. The language supports both whole-series and subsequence matching.}

{\textcolor{TSseekCurrentRevision}{To address the first challenge of compatible feature extraction and indexing,} \textsc{TSseek} represents stored series as line segments and derives query search regions in the same two-dimensional space of sequence positions and numerical values. The segments approximate the original readings within a bounded error and carry slope and value-range metadata. They are indexed across database hosts by \textbf{\textsc{TSseek-X}}, a distributed spatial index.}

\textcolor{TSseekCurrentRevision}{To address the second challenge of exact query processing at scale, \textsc{TSseek} uses pruning rules that preserve every true match. It derives spatial search windows and segment-level predicates from the query's value, trend, and duration constraints, accounting for segmentation error and the positions permitted by variable-length patterns and wildcards. \textsc{TSseek} uses selectivity estimates from grid statistics to plan index lookups, runs the lookups in parallel, and checks candidate series against the full query using their original readings.}

\subsection{\textcolor{TSseekRevisionRoundTwo}{Contributions}}
\textcolor{TSseekCurrentRevision}{We make four contributions.}

\subsubsection{\textcolor{TSseekRevisionRoundTwo}{Expressive Time Series Pattern Queries}}
\textcolor{TSseekCurrentRevision}{We present a regular-expression-inspired query model for value, trend, and duration constraints, with matching semantics for whole series and subsequences.}

\subsubsection{\textcolor{TSseekRevisionRoundTwo}{Compatible Feature Extraction and Indexing Mechanism}}
\textcolor{TSseekCurrentRevision}{We represent stored series and query constraints in a common position--value space. \textbf{\textsc{TSseek-X}} uses this representation to serve both matching modes from one distributed index over error-bounded line segments.}

\subsubsection{\textcolor{TSseekRevisionRoundTwo}{Exact Distributed Query Processing}}
\textcolor{TSseekCurrentRevision}{We derive conditions for safely ruling out series using their indexed segments. We show that refining the remaining candidates against their original readings returns all and only the matching series.}

\subsubsection{Experimental Evaluation}
Experiments on four real-world and benchmark datasets demonstrate \textsc{TSseek}'s scalability to 200 million time series with efficient response times
compared to baselines while returning exact matches. Ablation studies examine the effects of the segmentation threshold, slope-based pruning, and time-series length.

The paper is organized as follows. In  
Sec.~\ref{sec:related_work}, we present an overview of related work and its limitations. Sec.~\ref{sec:Prelim} introduces \textcolor{TSseekMotivationProposal}{the data model and  query constructs}. In \textcolor{TSseekMotivationProposal}{Secs.}~\ref{sec:overview} and~\ref{sec:Index}, we present  \textsc{TSseek}'s \textcolor{TSseekMotivationProposal}{architecture} and index structure, respectively. 
Query processing  is presented in Sec.~\ref{sec:QProcessing}. 
Finally, the experimental evaluation and 
concluding remarks are  in
Secs.~\ref{sec:exp} and~\ref{sec:conclusion}, respectively.

\begingroup
\section{Related Work}
\label{sec:related_work}
\begingroup
\color{TSseekRevisionRoundTwo}
\subsection{Time-Series Search}
\subsubsection{Search Task and Matching Scope}
Similarity search compares a target time series with stored series to find similar ones~\cite{ding2008querying}. In terms of matching scope, whole matching compares the target series with each of the time series stored in the database~\cite{agrawal1995fast, KeoghRatanamahatana2005ExactDTW, ShiehKeogh2009}, while subsequence matching compares it with a contiguous window within a series~\cite{8731498, linardi2020scalabledataseriessubsequence, chatzigeorgakidis2021twinsubsequencesearchtime, 10.1007/s10618-024-01005-2}.

\subsubsection{Exact vs. Approximate Similarity Search}
Independent of  whole versus  subsequence classification, similarity search techniques can also be distinguished as either {\em exact} or {\em approximate}. In exact search, the result set is guaranteed to satisfy the \textcolor{TSseekCurrentRevision}{query} and associated criteria (e.g., exact match or k-nearest neighbors)~\cite{KeoghRatanamahatana2005ExactDTW}. In contrast, approximate search may yield results that include false positives or false negatives relative to the exact result set, typically in exchange for indexing efficiency and faster query processing~\cite{SectionMin-HashApproximatingTSSearch, malkov2018efficientrobustapproximatenearest, 10.14778/3476249.3476255}. \textsc{TSseek} performs exact pattern retrieval, returning all and only series satisfying the specified query constraints.

\subsubsection{Distance-Based Similarity Search}
Whether retrieval is exact or approximate, distance-based methods for whole and subsequence matching commonly take a numerical sequence as the query~\cite{faloutsos1994fast, moon2001duality, SubsequenceMatchingOnStructuredTSData}. Faloutsos et~al.~\cite{faloutsos1994fast}, for example, discard a candidate if a lower bound on its distance to the query exceeds the search threshold. They then compute the actual distance for each remaining candidate using the original readings. For subsequence similarity search, MASS~\cite{10.1007/s10618-024-01005-2} computes a distance profile from a target time series to every equal-length subsequence, enabling nearest-neighbor or thresholded retrieval. SEAnet~\cite{wang2021seanet} learns autoencoder-based summaries for approximate similarity search.

\subsubsection{Pattern-Based Search}
Pattern-based approaches describe the desired behavior through rules, symbolic patterns, inter-series relations or natural-language requests~\cite{StockTSPatternMatching, CompositePatternMatchingInTS, Saxregex, Liu2024RelaQ, tan2026sonarts}. For example, Fu et~al.~\cite{StockTSPatternMatching} study template- and rule-based matching of financial patterns, while SAXRegEx~\cite{Saxregex} matches symbolic patterns in multivariate automotive sensor data. RelaQ~\cite{Liu2024RelaQ} also addresses pattern-based time-series querying through a visual GUI for analyst-driven exploration of inter-series relations (correlation, causality, lag). Sonar-TS~\cite{tan2026sonarts} answers natural-language requests by generating SQL searches over a feature index and Python programs to verify candidate windows against raw data. The closest system to our subsequence pipeline strategy is T-ReX~\cite{T_Rex}, a segment-based subsequence pattern search engine that allows users to express pattern queries through segment variables and compositional operators. Its query-time optimizations operate within each input series, without a persistent \textcolor{TSseekCurrentRevision}{dataset-level} index. \textsc{TSseek} is a pattern-based search system that evaluates explicit value, trend and duration constraints with exact, scalable processing over distributed \textcolor{TSseekCurrentRevision}{datasets}. Its segment index prunes candidate series before raw-data refinement returns all and only series satisfying the full query.
\par
\subsection{Complex Event Processing}
\textcolor{TSseekMotivationProposal}{\textcolor{TSseekRevisionRoundTwo}{Complex event processing (CEP)} systems combine predicates with ordered and repeated event patterns, as illustrated by SASE/SASE+~\cite{wu2006sase,gyllstrom2008kleene}, sequenced-event-set matching~\cite{cadonna2011sequenced} and ZStream~\cite{mei2009zstream}. DejaVu~\cite{dindar2009dejavu} supports live and archived event streams, while CORE~\cite{bucchi2022core} provides automaton-based recognition with time windows and partition-by correlations. For historical data, row-pattern recognition \textcolor{TSseekRevisionRoundTwo}{finds ordered sequences of rows satisfying user-defined predicates. SQL \texttt{MATCH\_RECOGNIZE} supports this task and} has been accelerated by prefiltering the stored events with indexes~\cite{korber2021index} or with joins chosen by a cost model~\cite{zhu2023joinnfa} before automaton evaluation; both filter raw events, whereas \textsc{TSseek} derives necessary conditions on an error-bounded segment approximation shared across queries.}

\subsection{Feature Extraction \& Indexing Techniques}
\textcolor{black}{Many feature extraction techniques have been proposed in the literature, including
DFT~\cite{faloutsos1994fast},
DWT~\cite{chan1999efficient},
PAA~\cite{keogh2001dimensionality},
SAX~\cite{lin2007experiencing},
and iSAX~\cite{ShiehKeogh2009}.} Numerical summaries can be indexed using traditional Spatial Access Methods (SAMs) such as the R-tree~\cite{guttman1984r} or its variants~\cite{arge2008priority,beckmann1990r,kamel1993hilbert}. Custom-designed indexes include the iSAX Binary Tree~\cite{Camerra2010iSAX2} over iSAX words and {\em sigTree}~\cite{zhang2019tardis} over transposed iSAX signatures.

Another class of methods learns representations from data. \textcolor{TSseekMotivationProposal}{SEAnet~\cite{wang2021seanet} learns autoencoder-based summaries \textcolor{TSseekRevisionRoundTwo}{that replace PAA before SAX symbolization and iSAX indexing}, while Series2Vec~\cite{foumani2024series2vec} learns self-supervised representations in the time and frequency domains for classification.} \textcolor{TSseekMotivationProposal}{\textsc{TSseek} instead indexes \textcolor{TSseekRevisionRoundTwo}{error-bounded segment geometries in position--value space using \textbf{\textsc{TSseek-X}}} and derives spatial probes from the query's constraints (\textcolor{TSseekRevisionRoundTwo}{Secs.~\ref{sec:Index}--\ref{sec:QProcessing}}).}

\subsection{Time Series Database Systems}
\textcolor{TSseekMotivationProposal}{Time-series databases~\cite{influxdb} such as Apache IoTDB~\cite{wang2020apacheiotdb} and InfluxDB support storage, ingestion and analytical queries.} IoTDB's \texttt{MATCH\_RECOGNIZE} evaluates numerical pattern queries over sequences of input rows~\cite{iotdbpatternquery}. \textsc{TSseek} indexes error-bounded segments to prune candidate series before exact refinement.
\par\endgroup
\endgroup

\section{\textcolor{TSseekMotivationProposal}{\textsc{TSseek} Query and Data Model}}
\label{sec:Prelim}

\label{sec:definitions}

\begingroup\color{black}

\noindent \textbf{Definition 1. Time Series:} A time series \( X =\textcolor{TSseekMotivationProposal}{\langle}x_1, x_2, \dots, x_m\textcolor{TSseekMotivationProposal}{\rangle} \), where \( x_i \in \mathbb{R} \) for all \( 1 \leq i \leq m \), is an ordered sequence of \( m \) real-valued variables. \textcolor{TSseekRevisionRoundTwo}{Readings are sampled at a fixed interval, so timestamps are implicit.}

\noindent \textbf{Definition 2. Time Series Dataset:} A time series dataset \( \mathcal{T} = \{X_1, X_2, \dots, X_{\textcolor{TSseekMotivationProposal}{N}}\} \) is a \textcolor{TSseekCurrentRevision}{set} of \( N \) time series objects \( X_i \), each of length $m$.

\textcolor{TSseekMotivationProposal}{\textbf{\textcolor{TSseekRevisionRoundTwo}{Dataset Scope.}} \textcolor{TSseekRevisionRoundTwo}{The equal-length records in Def.~2 use the same sampling interval, so whole matching compares records covering the same duration.} DSTree~\cite{wang2013data} and \textsc{TARDIS}~\cite{zhang2019tardis} also assume equal-length series for whole matching. \textcolor{TSseekRevisionRoundTwo}{Pattern and subsequence lengths may still vary.} 
\eltabakhcomment{}
\textcolor{TSseekRevisionRoundTwo}{\textsc{TSseek} evaluates} query predicates on the stored readings \textcolor{TSseekRevisionRoundTwo}{without applying normalization}. Sec.~\textcolor{TSseekRevisionRoundTwo}{\ref{sec:features}} discusses normalization \textcolor{TSseekRevisionRoundTwo}{in more detail}.}

\begingroup\color{TSseekCurrentRevision}
\noindent\textbf{Definition 3. \textsc{TSseek} Query:} A \textsc{TSseek} query $Q=\langle q_1,\ldots,q_k\rangle$, $k\geq1$, consists of numerical constructs evaluated under whole-matching or subsequence-matching semantics. Each construct has one of the following forms:
\[
q_i ::= S(v)\mid R(I)\mid P(D,\delta,L,\sigma)\mid W(L).
\]
Table~\ref{review-original:tab:query-constructs} explains the notation and parameters for Singleton ($S$), Range ($R$), Pattern ($P$), and Wildcard ($W$).

\begin{table*}[t]
\centering
\color{TSseekCurrentRevision}
\caption{\textcolor{TSseekCurrentRevision}{\textsc{TSseek} Query Constructs}}
\label{review-original:tab:query-constructs}
\footnotesize
\setlength{\tabcolsep}{4pt}
\renewcommand{\arraystretch}{1.04}
\begin{tabularx}{\textwidth}{@{}>{\raggedright\arraybackslash}p{0.14\textwidth}>{\raggedright\arraybackslash}p{0.20\textwidth}>{\raggedright\arraybackslash}X@{}}
\toprule
\textbf{Construct} & \textbf{Notation} & \textbf{Meaning and parameters} \\
\midrule
\textbf{Singleton (S)} & $v$ & Fixes one reading to the numerical value $v\in\mathbb{R}$. \\
\addlinespace
\textbf{Range (R)} & An interval, e.g., $[v_1,v_2)$ & Restricts one reading to the value interval $I$. Each endpoint may be inclusive or exclusive. \\
\addlinespace
\textbf{Pattern (P)} & $\big\lvert\textcolor{blue}{D}\,\frac{\textcolor{red}{\delta}}{\textcolor{red}{\sigma}}\,(L)\big\rvert$ & Every reading in the block must lie in the value domain $D$. The direction $\delta$ is increasing ($+$), decreasing ($-$), flat ($=$), or unspecified ($?$). A single integer $L$ fixes the block length to $L$ readings. An interval specifies the permitted integer lengths of a variable-length block. The optional magnitude $\sigma$ constrains the change between consecutive readings within the block. It can be a fixed value or an interval, in absolute or relative terms. \\
\addlinespace
\textbf{Wildcard (W)} & $\textcolor{red}{W}$, $*$, or $(*,L)$ & Allows any value at the positions it occupies. Its length $L$ can be fixed or variable. A trailing $W$ in a whole-matching query consumes the remaining readings. \\
\bottomrule
\end{tabularx}
\end{table*}

Whole-matching queries combine constructs in query order. A sequence of Singletons specifies an exact numerical sequence. Subsequence queries use the same Pattern construct to specify one or more matching windows, with constraints on their starts, ordering and gaps. A Range condition is represented by a length-one Pattern with the same value interval, unspecified direction, and no magnitude constraint. A Singleton is the special case whose domain is $[v,v]$. Definition~4 specifies when a series matches a query under each mode.

\textbf{Transition Tolerance.}
An optional absolute \texttt{Threshold} $\tau$ (default $0$) relaxes consecutive-reading predicates: a flat Pattern permits $|x_{j+1}-x_j|\le\tau$, and step bounds widen by $\tau$. Comparisons also allow floating-point roundoff. The tolerance never widens a value domain or permits a direction reversal. Subsequence queries additionally support percentage tolerances, computed from the absolute value of the specified step, the midpoint of a step interval, or the preceding reading for a flat Pattern. Thus, $\tau$ is part of the query's meaning and can change its answers. By comparison, $\varepsilon$ bounds segmentation error and expands index probes without relaxing the final-match predicates.

\noindent\textbf{Definition 4. \textsc{TSseek} Query Placement and Match:} Let $X=\langle x_1,\ldots,x_m\rangle$. A placement of $Q$ in $X$ assigns each construct $q_i$ a block $X[s_i:e_i]$, where $s_i$ and $e_i$ denote its starting and ending reading positions. The matcher determines these positions subject to the query's length and placement constraints. The block's length $e_i-s_i+1$ must be permitted by the construct, and its readings must satisfy the construct's predicates. Nonempty blocks have integer endpoints $1\leq s_i\leq e_i\leq m$. A zero-length remainder Wildcard has $e_i=s_i-1$ and consumes no readings.

\textbf{Whole matching} requires $s_1=1$, $e_k=m$, and $s_{i+1}=e_i+1$ for $1\leq i<k$. Thus, each construct starts immediately after the preceding one ends; their blocks do not overlap and together cover the entire series.

\textbf{Subsequence matching} assigns each Pattern a contiguous window. Window starts are free unless fixed or bounded by the query. For ordered matching, windows follow query order, and each gap $g_i=s_{i+1}-e_i-1$ must satisfy its specified bounds: adjacency requires $g_i=0$, while a flexible gap permits any $g_i\geq0$. Unordered matching permits any temporal order with non-overlapping windows and flexible gaps. Readings outside the selected windows are unconstrained.

A series matches $Q$ if at least one placement satisfies these requirements.
\par\endgroup

\additionalinfo{}

\additionalinfo{}

\par\endgroup

\subsection{\textcolor{TSseekCurrentRevision}{Query Examples and SQL-like Notation}}
\label{sec:Lang}

\begingroup\color{black}

\textcolor{TSseekRevisionRoundTwo}{Regular expressions describe patterns over sequences of symbols and are used in text search~\cite{Thompson1968,Clarke1995} and search over symbolic time series representations~\cite{Saxregex,tan2026sonarts}. Inspired by their syntax, \textsc{TSseek} incorporates four constructs for expressing numerical query constraints. These constraints are evaluated by \textsc{TSseek}'s query processor rather than a regular-expression matching engine. }\textcolor{TSseekCurrentRevision}{The following examples show how these constructs specify whole-matching and subsequence queries, first in the notation of Table~\ref{review-original:tab:query-constructs} and then in SQL-like form.}

\par\endgroup

\begingroup\color{TSseekMotivationProposal}
\textbf{Example \textcolor{TSseekMotivationProposal}{1}. Household-Power Pattern Query.}
\textcolor{TSseekRevisionRoundTwo}{We instantiate the motivating whole-matching query with illustrative numerical settings and express it using the constructs in Table~\ref{review-original:tab:query-constructs}. We also give a separate subsequence query for rapid demand increases.} Values are in watts, and lengths count readings. For illustration, consider records of $m=16$ readings.

\textbf{Whole-Matching Form.}
\par\nopagebreak[4]\begingroup
\smallskipamount=3pt\relax
\smallskip\noindent
\sbox{\TSseekWholeFormulaBox}{$\displaystyle\TSseekWholeRunningQuery$}
\pgfmathsetmacro{\TSseekQueryScale}{\the\columnwidth/\the\wd\TSseekWholeFormulaBox}
\scalebox{\TSseekQueryScale}{\usebox{\TSseekWholeFormulaBox}}\par\smallskip\endgroup
Here, (a) constrains the first reading; (b) covers the next six readings with internal increases of 100--200\,W; and (c) specifies five to seven decreasing readings without a step size. Both Pattern domains apply to every reading in their respective blocks. If (c) consumes $\ell\in\{5,6,7\}$ readings, the remainder Wildcard (d) consumes $16-1-6-\ell=9-\ell$ readings: four, three, or two, respectively.

\textbf{Subsequence-Matching Query.}
To screen for rapid demand increases without requiring the complete rise-and-decline profile, the analyst searches for the increasing Pattern (b) anywhere in a record:
\par\nopagebreak[4]\begingroup
\smallskipamount=3pt\relax
\sbox{\TSseekWholeFormulaBox}{$\displaystyle\TSseekWholeRunningQuery$}
\pgfmathsetmacro{\TSseekQueryScale}{\the\columnwidth/\the\wd\TSseekWholeFormulaBox}
\smallskip\noindent\makebox[\columnwidth][c]{%
\scalebox{\TSseekQueryScale}{$\displaystyle
\color{black}
Q_{\mathrm{sub}}=\big\langle
\underbrace{\big|\textcolor{blue}{[200,800]}\,\frac{\textcolor{red}{+}}{\textcolor{red}{[100,200]}}\,(6)\big|}_{\text{(b)}}
\big\rangle .
$}%
}\par\smallskip\endgroup
A match consists of six consecutive readings in 200--800\,W, with each of the five internal steps increasing by 100--200\,W. No preceding low reading or subsequent decline is required. Readings outside the matched span are unconstrained.

\textcolor{TSseekCurrentRevision}{\textbf{SQL-like Query Notation.}}
\textcolor{TSseekCurrentRevision}{\textsc{TSseek} accepts queries as compact construct expressions. For readability, we present query examples in SQL-like notation to make the target dataset, matching mode and numerical constraints explicit. This notation has the following form:}

\begin{lstlisting}[
    language={},
    basicstyle=\ttfamily\footnotesize\color{TSseekRevisionRoundTwo},
    breaklines=true,
    columns=fullflexible,
    keepspaces=true,
    aboveskip=3pt,
    belowskip=3pt
]
<query> ::= SELECT * FROM <dataset>
            WHERE <mode>(<constructs>);
<mode> ::= MATCHES | MATCHES_SUBSEQUENCE
<constructs> ::= <construct>
               | <construct>, <constructs>
\end{lstlisting}
\textcolor{TSseekRevisionRoundTwo}{Here, \texttt{<dataset>} identifies the \textcolor{TSseekCurrentRevision}{time series dataset $\mathcal{T}$}, and \texttt{<construct>} is one of the four types in Table~\ref{review-original:tab:query-constructs}. \texttt{PATTERN} uses \texttt{DOM}, \texttt{DIR} and \texttt{LEN}, with optional \texttt{MAG}.}

\textbf{Example \textcolor{TSseekMotivationProposal}{2}.}
The listings express both queries from Example~1 in SQL-like notation. Omitting \texttt{MAG} leaves the decreasing Pattern's step size unspecified.

\par\vspace{3pt}\noindent
\begin{minipage}[t]{0.48\linewidth}
\vspace{0pt}
\begin{lstlisting}[
    language=SQL,
    basicstyle=\ttfamily\footnotesize\color{TSseekMotivationProposal},
    keywordstyle=\bfseries\color{TSseekMotivationProposal},
    commentstyle=\color{TSseekMotivationProposal},
    stringstyle=\color{TSseekMotivationProposal},
    breaklines=true,
    columns=fullflexible,
    keepspaces=true,
    aboveskip=0pt,
    belowskip=0pt
]
-- Whole matching
SELECT *
FROM <Power records>
WHERE MATCHES (
  RANGE([100,200]),
  PATTERN(DOM=[200,800],
    DIR='+', MAG=[100,200],
    LEN=6),
  PATTERN(DOM=[200,800],
    DIR='-', LEN=[5,7]),
  WC([*])
);
\end{lstlisting}
\end{minipage}\hfill
\begin{minipage}[t]{0.48\linewidth}
\vspace{0pt}
\begin{lstlisting}[
    language=SQL,
    basicstyle=\ttfamily\footnotesize\color{TSseekMotivationProposal},
    keywordstyle=\bfseries\color{TSseekMotivationProposal},
    commentstyle=\color{TSseekMotivationProposal},
    stringstyle=\color{TSseekMotivationProposal},
    breaklines=true,
    columns=fullflexible,
    keepspaces=true,
    aboveskip=0pt,
    belowskip=0pt
]
-- Subsequence matching
SELECT *
FROM <Power records>
WHERE MATCHES_SUBSEQUENCE (
  PATTERN(DOM=[200,800],
    DIR='+', MAG=[100,200],
    LEN=6)
);
\end{lstlisting}
\end{minipage}\par\vspace{3pt}

\par\endgroup

\begingroup\color{black}
\noindent \textbf{Problem Statement:} 
\textcolor{TSseekMotivationProposal}{Given a \textsc{TSseek} \textcolor{TSseekCurrentRevision}{query} $Q$, a matching mode (whole or subsequence), and a time series dataset $\mathcal{T}$, return all and only the series $X\in\mathcal{T}$ having a match that satisfies every constraint of $Q$ under that mode.}

\par\endgroup

\section{System Overview}
\label{sec:overview}
\label{sec:Architecture}

\begin{figure}[H]
 \centering
 \includegraphics[width=\columnwidth]{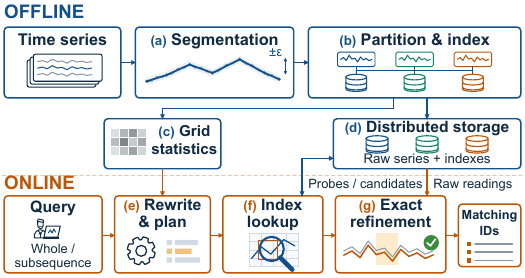}
 
 \caption{\textsc{TSseek} System Architecture. \textcolor{TSseekRevisionRoundTwo}{Letters identify blocks.}}
 \label{fig:Arch}
 
\end{figure}

\begingroup\color{TSseekCurrentRevision}
Fig.~\ref{fig:Arch} summarizes \textsc{TSseek}'s offline and online processing phases. During \textbf{offline processing (blocks~(a)--(d))}, \textsc{TSseek} extracts $\varepsilon$-bounded line segments from raw series, partitions raw series and segments for distributed storage, and builds \textbf{\textsc{TSseek-X}}, a distributed spatial index over the segments. Grid statistics are aggregated during index construction. During \textbf{online processing (blocks~(e)--(g))}, queries are rewritten and planned before execution. Parallel index lookups identify candidate series while preserving every true match. Refinement retrieves their original readings, checks the complete original query, and returns matching series IDs.

Whole and subsequence matching reuse this index and filter-then-refine pipeline. Grid statistics guide index lookup planning in both modes. Whole matching selects the probes of the construct estimated to be most selective. For subsequence patterns with fixed or bounded starts, the statistics guide a choice between position-based spatial probes and a value-range condition over all positions. Free-start and relative-position patterns use value-range conditions; multi-pattern queries combine the conditions of all sub-patterns. Refinement enforces the corresponding placement semantics of Definition~4. Sections~\ref{sec:Index} and~\ref{sec:QProcessing} detail indexing and query processing, respectively.
\par\endgroup

\section{\textsc{TSseek} Indexing Framework}
\label{sec:Index}

\begingroup\color{TSseekCurrentRevision}
To support indexed pattern search, \textsc{TSseek} represents stored series and query conditions in a common position--value space. Stored series are approximated by error-bounded line segments, while the query's value and position requirements define spatial search regions. Our distributed spatial index, \textbf{\textsc{TSseek-X}}, uses this representation to identify candidate series for exact refinement while preserving every true match. Sections~\ref{sec:features} and~\ref{sec:index} describe feature extraction and the distributed index structure, respectively.
\par\endgroup

\subsection{\textsc{TSseek} Feature Extraction}
\label{sec:features}

\begingroup\color{TSseekMotivationProposal}
\textbf{Key Insight.}
A close examination of the supported query constructs (e.g., Singletons, Ranges, and Patterns) reveals that they primarily constrain the search space to specific regions within a two-dimensional domain, where the x-axis corresponds to the temporal dimension (equivalent to the sequence position), while the y-axis corresponds to the value dimension.
For instance, consider the whole-matching household-power query from Example~\textcolor{TSseekMotivationProposal}{1} over a 16-reading record:
\par\smallskip\noindent
\sbox{\TSseekWholeFormulaBox}{$\displaystyle\TSseekWholeRunningQuery$}
\pgfmathsetmacro{\TSseekQueryScale}{\the\columnwidth/\the\wd\TSseekWholeFormulaBox}
\scalebox{\TSseekQueryScale}{\usebox{\TSseekWholeFormulaBox}}\par\smallskip
\begingroup\color{TSseekCurrentRevision}
In the query formula above, labels~(a)--(d) identify the initial Range, fixed-length rise, variable-length decline, and remainder Wildcard, respectively. Fig.~\ref{fig:QRectangles} maps their value bounds and permitted positions into two-dimensional regions. Range~(a) restricts reading~1 to 100--200\,W, while Pattern~(b) restricts readings~2--7 to 200--800\,W. Pattern~(c) begins at reading~8 and uses the same 200--800\,W domain, but permits five to seven readings. Its possible extent is divided into three regions: $c_1$ covers the minimum five readings, 8--12, while $c_2$ and $c_3$ extend it to readings~13 and~14. Thus, decline lengths five, six and seven occupy $c_1$, $c_1\cup c_2$, and $c_1\cup c_2\cup c_3$, respectively.

In Fig.~\ref{fig:QRectangles}, $PD$ denotes the variable-length Pattern~(c), while $e$ and $s$ mark ending and starting positions. The three length choices give decline endings at readings~12, 13 and~14, labeled $PD_{e1}$--$PD_{e3}$. The corresponding Wildcard starts are at readings~13, 14 and~15, labeled $W_{s1}$--$W_{s3}$. Regions $d_1,d_2,d_3$ subdivide the Wildcard's possible extent. It occupies $d_1\cup d_2\cup d_3$, $d_2\cup d_3$, or $d_3$, respectively, with unrestricted values through reading~16. Overlapping shading represents these alternative placements; within each chosen placement, the constructs remain adjacent and non-overlapping.
\par\endgroup

\textcolor{TSseekCurrentRevision}{The example shows how value bounds and permitted construct lengths define query regions in position--value space. To enable spatial index lookups using these regions, we approximate stored time series as sequences of error-bounded two-dimensional line segments.}

This ensures that both the dataset objects and the query representations are mapped into the same position--value feature space. Consequently, spatial index structures can be employed to accelerate query processing. Sec.~\ref{sec:QProcessing} explains how query regions yield index probes while accounting for segmentation error and uncertain boundaries, followed by refinement against the original readings.

\begin{figure}[t]
 \color{TSseekMotivationProposal}

 \centering
 \includegraphics[width=\linewidth]{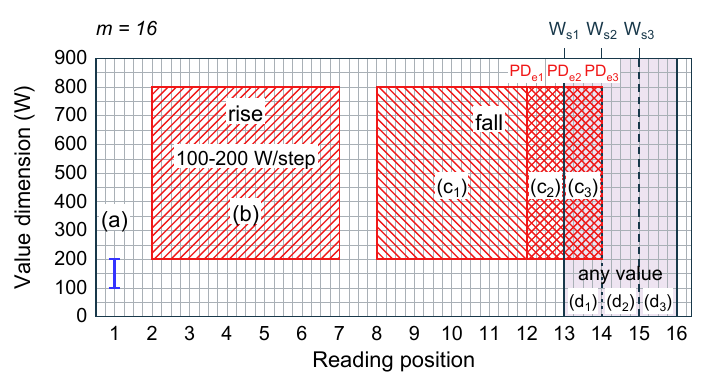}
 \caption{Two-dimensional visualization for an illustrative \textcolor{TSseekCurrentRevision}{query}.}
 \label{fig:QRectangles}
\end{figure}
\par\endgroup

\begingroup\color{TSseekMotivationProposal}
\textbf{Normalization.}
\textsc{TSseek} evaluates predicates in the original value domain, retaining application-specific levels and additive changes, as in numerical pattern queries~\cite{T_Rex}. Example~\textcolor{TSseekMotivationProposal}{1}'s subsequence query accepts $\langle200,300,400,500,600,800\rangle$\,W but rejects $\langle600,900,1200,1500,1800,2400\rangle$\,W: the latter violates its value and step bounds, although both six-reading windows have identical z-normalized shapes. \textcolor{TSseekCurrentRevision}{Independently z-normalizing each candidate window, as in normalized shape search~\cite{Rakthanmanon2012}, would require converting the query's value and step bounds using that window's own mean and standard deviation to preserve the specified constraints in watts.}
The current system does not perform this independent per-window normalization. A common fixed normalization can preserve the example's value and additive-step constraints if both the query bounds and absolute tolerances are transformed
consistently. We do not provide per-window offset/scale invariance.
\par\endgroup

\begingroup\color{TSseekMotivationProposal}
\textbf{Segmentation.}
\begingroup\color{TSseekCurrentRevision}
\textsc{TSseek} approximates the stored readings with a fixed-slope variant of piecewise linear approximation (PLA)~\cite{KeoghICDM2001,KeoghSurvey2004}, summarized in Algorithm~\ref{alg:piecewise_approx} (Fig.~\ref{fig:Arch}, block~(a)). The first two readings define a segment's line, which remains fixed as the segment grows. For each subsequent reading, the algorithm computes $E$, the absolute difference between its value and the line's value at the same reading position. If $E<\varepsilon$, the segment extends along this line to the new reading position. Its new endpoint is the reading's vertical projection onto the line.

If $E\geq\varepsilon$, the current segment is finalized at its existing endpoint. The rejected reading and its successor initialize the next segment's line. If the rejected reading is the final reading, a terminal segment instead joins the last two original readings. Both segments then cover the penultimate reading position.

The error bound supports spatial candidate preservation in Sec.~\ref{sec:selectivity}; Sec.~\ref{sec:DataDriven} describes threshold configuration.

\textbf{Complexity.} Algorithm~\ref{alg:piecewise_approx} performs constant work per reading and takes $O(m)$ time per series.
\par\endgroup

\par\endgroup

\begingroup\color{TSseekMotivationProposal}
\begin{algorithm}[htbp]
 \color{TSseekMotivationProposal}\small
 \DontPrintSemicolon
 \KwIn{Reading points $X_j=(j,x_j)$, $1\leq j\leq m$, $m\geq2$; threshold $\varepsilon>0$}
 \KwOut{List of line segments $L_X$}
 \Begin{
   // Initialization\\
   Construct initial segment $S_1$ by connecting $X_1$ and $X_2$\;
   Set current segment index $i\gets1$\;
   Initialize $j\gets3$ // The third reading\\
   \While{$j\leq m$}{
     Let $\ell_i$ be the line of $S_i$ and $P_j=(j,\ell_i(j))$\;
     // Vertical error at the same reading position\\
     $E\gets|x_j-\ell_i(j)|$\;
     \uIf{$E<\varepsilon$}{
       Extend $S_i$ to $P_j$ // maintains the same slope\\
     }
     \Else{
       Finalize current segment $S_i$\;
       \uIf{$j<m$}{
         Construct $S_{i+1}$ by connecting $X_j$ and $X_{j+1}$\;
         $j\gets j+1$ // skip the next defining reading\\
       }
       \Else{
         Construct terminal $S_{i+1}$ through $X_{j-1}$ and $X_j$\;
       }
       $i\gets i+1$\;
     }
     $j\gets j+1$\;
   }
   \Return{all segments $S_i$ in $L_X$}\;
 }
 \caption{Fixed-Slope Time Series Segmentation.}
 \label{alg:piecewise_approx}
\end{algorithm}
\par\endgroup

\subsection{\textsc{TSseek} Index Structure}
\label{sec:index}

\begingroup\color{TSseekMotivationProposal}
\textbf{Distributed Storage.}
We use Spark for fault-tolerant parallel execution of segmentation, statistics collection, and segment redistribution~\textcolor{black}{\cite{Zaharia2016SparkCACM}}. PostgreSQL stores each raw time series as a single tuple for retrieval by series ID lookup. \textcolor{TSseekCurrentRevision}{The corresponding segments are stored as separate rows in segment tables, with each row retaining the originating series ID.} Further,
PostGIS indexes the segments' \texttt{LINESTRING} geometries using GiST-based R-trees, which persist across queries~{\cite{postgis2024}}. SQL probes identify promising candidate IDs in the \textsc{TSseek} index lookup stage. Spark tasks then retrieve their raw readings and perform the subsequent exact refinement step.
Multiple partitions per host permit concurrent SQL probes against separate 
segment tables and their indexes (Sec.~\ref{sec:distExec}).
\par\endgroup

\begingroup\color{TSseekCurrentRevision}
\textbf{Distributed Spatial Index (\textsc{TSseek-X}).} \textsc{TSseek-X} is a distributed spatial index over the line segments extracted from each stored time series. The construction process in Algorithm~\ref{alg:distributed_index} (Fig.~\ref{fig:Arch}, blocks~(b) and~(d)) organizes the dataset into $N_{\mathrm{tbl}}$ logical partitions, assigning each time series to exactly one partition. Spark workers apply Algorithm~\ref{alg:piecewise_approx} to the series concurrently, retaining their assigned partition identifiers. All segments from the same logical partition are stored together in one segment table. Each segment table is paired with a raw-series table through their shared partition identifier. Using that identifier and the series ID, refinement retrieves each candidate's original readings from the corresponding raw-series table.
\par\endgroup

\begin{algorithm}[htbp]
 \small
 \DontPrintSemicolon
 \KwIn{Time series dataset $\textcolor{TSseekMotivationProposal}{\mathcal{T}}$ stored in HDFS, threshold $\varepsilon$}
 \KwOut{Distributed spatial index on cluster machines}
 \Begin{
   // Segmentation and Distribution\\
   \ForEach{time series $X$ in dataset $\textcolor{TSseekMotivationProposal}{\mathcal{T}}$}{
     \textcolor{TSseekCurrentRevision}{$L_X$ $\leftarrow$ Algorithm~\ref{alg:piecewise_approx} ($X$, $\varepsilon$)}\;
     \textcolor{TSseekMotivationProposal}{Let $P$ be $X$'s assigned logical partition}\;
     Distribute segments $L_X$ to partition $P$ across the cluster\;
   }
   // Local Index Construction\\
   \ForEach{partition $P$}{
     \textcolor{TSseekMotivationProposal}{Create PostgreSQL table $T_P$ with a \texttt{LINESTRING} geometry column and a GiST R-tree index}\nllabel{ln:index_create}\;
     \textcolor{TSseekMotivationProposal}{Bulk-insert the segments of $P$ into $T_P$}\nllabel{ln:index_load}\;
   }
   Store partition metadata for query routing\;
 }
 \caption{ Spatial Index Construction.}
 \label{alg:distributed_index}
\end{algorithm}

\textcolor{TSseekMotivationProposal}{For each partition, its PostgreSQL segment table and corresponding spatial index are created before bulk insertion (Alg.~\ref{alg:distributed_index}, lines~\ref{ln:index_create}--\ref{ln:index_load}).}

\begingroup\color{TSseekCurrentRevision}
\textbf{Data Statistics.} During index construction (Fig.~\ref{fig:Arch}, block~(c)), \textsc{TSseek} partitions the position--value space into a regular grid. Spark workers use the same grid to count, within each data partition, the segments intersecting each cell, including segments that only touch the cell boundary. These partition counts are summed into one count per cell across the complete indexed dataset and saved for query processing. The resulting statistics guide index lookup planning (Sec.~\ref{sec:QProcessing}); Sec.~\ref{sec:DataDriven} describes how the grid geometry is configured.
\par\endgroup

\section{Query Processing}
\label{sec:QProcessing}

\textcolor{TSseekCurrentRevision}{\textsc{TSseek} returns the IDs of the time series that match a query $Q$ under the semantics of Definition~4. Query planning rewrites and simplifies the query and uses selectivity estimates to guide index lookup. Distributed execution has two stages. Index lookup identifies candidate series. Exact refinement retrieves their original readings and checks them against the full query. Sec.~\ref{sec:SubseqProcessing} describes how index lookup and refinement are specialized for subsequence matching.}

\subsection{Query Rewriting and Simplification}
\label{sec:QRewriting}
\begingroup\color{black}
\def\TSseekRewriteChange#1{\textcolor{TSseekMotivationProposal}{#1}}
\def\TSseekRewriteTrend#1#2#3{\big|\textcolor{blue}{#1}\,\overset{\textcolor{red}{#2}}{\text{---}}\,(#3)\big|}
\def\TSseekRewritePattern#1#2#3#4{\big|\textcolor{blue}{#1}\,\frac{\textcolor{red}{#2}}{\textcolor{red}{#3}}\,(#4)\big|}
\def\TSseekRewriteDisplay#1{\par\smallskip\noindent{\color{black}$\displaystyle\begin{aligned}#1\end{aligned}$}\par\smallskip}
\def\TSseekLengthDisplay#1{\par\smallskip\noindent\begingroup\color{black}\sbox{\TSseekWholeFormulaBox}{$\displaystyle #1$}\ifdim\wd\TSseekWholeFormulaBox>\linewidth\resizebox{\linewidth}{!}{\usebox{\TSseekWholeFormulaBox}}\else\usebox{\TSseekWholeFormulaBox}\fi\endgroup\par\smallskip}
\textcolor{TSseekRevisionRoundTwo}{We employ the following rewriting and simplification rules (Fig.~\ref{fig:Arch}, block~(e)). The first two apply  to whole matching; the third rule on early termination due to conflicting constraints applies to both matching modes.} \TSseekRewriteChange{The length-rewriting examples below use 12-reading records; $Q_A$ is the query of Example~\textcolor{TSseekMotivationProposal}{1} evaluated at this shorter record length.}

\textbf{Resizing Variable-Length Patterns and Wildcards.}
Variable-length \TSseekRewriteChange{Patterns and Wildcards} can be resized---or, in some cases, eliminated---to enforce the whole-match constraint. \TSseekRewriteChange{After resolving any implicit remainder as a Wildcard, the construct lengths must sum to the stored series length. Consider the following household-power query:}
\TSseekLengthDisplay{
Q_A=\big\langle\textcolor{blue}{[100,200]},\TSseekRewritePattern{[200,800]}{+}{[100,200]}{6},\TSseekRewriteTrend{[200,800]}{-}{[5,7]},\textcolor{red}{W}\big\rangle .}
\TSseekRewriteChange{The initial Range and fixed-length rise consume $1+6=7$ readings, leaving five. Since the decreasing Pattern requires at least five readings, its length becomes five. Thus, the final Wildcard is eliminated:}
\TSseekLengthDisplay{
Q_A'=\big\langle\textcolor{blue}{[100,200]},\TSseekRewritePattern{[200,800]}{+}{[100,200]}{6},\TSseekRewriteTrend{[200,800]}{-}{5}\big\rangle .}
\TSseekRewriteChange{For a second query in the same application, let
us shorten the fixed-length rise to four readings, and follow the decline with a variable-length rise:}
\TSseekLengthDisplay{
Q_B=\big\langle\textcolor{blue}{[100,200]},\TSseekRewritePattern{[200,800]}{+}{[100,200]}{4},\TSseekRewriteTrend{[200,800]}{-}{[3,7]},\TSseekRewriteTrend{[300,600]}{+}{[2,6]}\big\rangle .}
\TSseekRewriteChange{The minimum length requirement is $1+4+3+2=10$, leaving two additional readings. The last two length intervals can therefore be tightened to $[3,5]$ and $[2,4]$, while retaining the condition that their chosen lengths must sum to seven:}
\TSseekLengthDisplay{
Q_B'=\big\langle\textcolor{blue}{[100,200]},\TSseekRewritePattern{[200,800]}{+}{[100,200]}{4},\TSseekRewriteTrend{[200,800]}{-}{[3,5]},\TSseekRewriteTrend{[300,600]}{+}{[2,4]}\big\rangle .}
\TSseekRewriteChange{These length rewrites preserve the exact answer set.}

\textbf{Superset Coverage of Overlapping Regions.}
\textcolor{TSseekCurrentRevision}{Variable-length Patterns can shift the boundary between constructs. If different constructs can occupy a reading position under different length choices, the filtering query uses a Range enclosing their value domains. The Range constrains the value at that position and imposes no direction or step-size constraint. If a Wildcard can occupy the position, the value remains unconstrained.}

\textcolor{TSseekCurrentRevision}{In $Q_B'$, the decline always covers readings~6--8, and the final rise always covers readings~11--12. Readings~9 and~10 can belong to either the decreasing Pattern with domain $[200,800]$ or the following increasing Pattern with domain $[300,600]$. The filtering query therefore uses a Range of $[200,800]$ at each of those two positions:}
\TSseekLengthDisplay{\begin{aligned}
&\textstyle Q_{B,\mathrm{filter}}=\big\langle\textcolor{blue}{[100,200]},\TSseekRewritePattern{[200,800]}{+}{[100,200]}{4},\\
&\quad\textstyle\TSseekRewriteTrend{[200,800]}{-}{3},\textcolor{blue}{[200,800]},\textcolor{blue}{[200,800]},\TSseekRewriteTrend{[300,600]}{+}{2}\big\rangle .
\end{aligned}}
\textcolor{TSseekCurrentRevision}{Every match to the original query satisfies the filtering query. Refinement checks the complete original query.}

\textbf{Early Termination Due to Conflicting Constraints.}
\textcolor{TSseekCurrentRevision}{\textsc{TSseek} can return an empty result before probing the index when the query constraints rule out every match. Consider an increasing Pattern of eight readings with a step interval of $[100,200]$\,W and a value domain of $[200,800]$\,W. With zero query tolerance ($\tau=0$), each step must meet the 100\,W minimum. Seven steps therefore require a rise of at least 700\,W, exceeding the 600\,W allowed by the value domain. Such checks account for the query's tolerance and interval endpoints. A query is also infeasible if its minimum required length exceeds the stored-series length. In whole matching, the permitted construct lengths must admit a placement that covers the entire series. If a whole-matching query ends with an explicit remainder Wildcard $W$, the preceding constructs may occupy fewer than $m$ readings because $W$ consumes the remaining readings.}

\begin{figure}[t]
 \centering
 \includegraphics[width=0.85\linewidth, height=7.5cm]{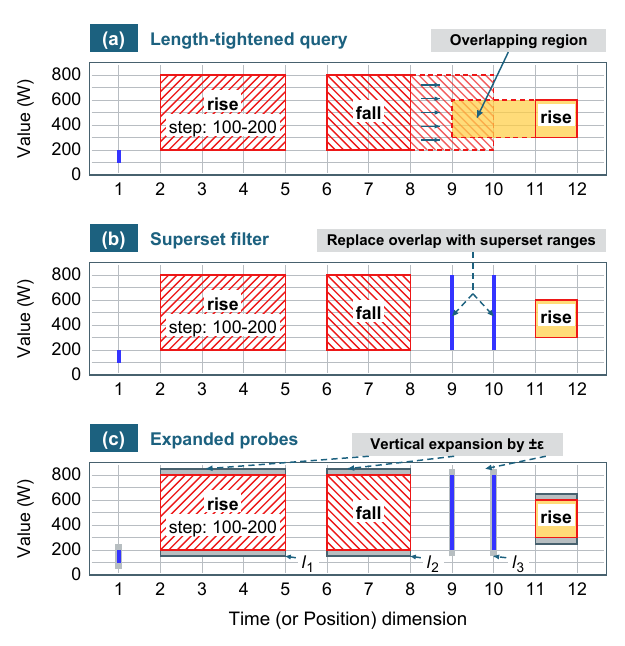}
 \caption{Example of query rewriting and expansion.}
 \label{fig:queryre-write}
\end{figure}
\par\endgroup

\subsection{Selectivity Estimation for Index Lookup}
\label{sec:selectivity}

\begingroup\color{TSseekMotivationProposal}
For a whole-matching query, \textsc{TSseek} builds the probe windows for every non-Wildcard construct and then uses grid statistics to select the one
construct for probing the index.

\textbf{Probe Windows.}
Each non-Wildcard construct maps to windows in the position--value space whose value bounds are expanded by $\pm\varepsilon$ to absorb the segmentation error (Fig.~\ref{fig:queryre-write}(c)).
Refinement still evaluates the original predicates. A Range $[v_{\min},v_{\max}]$ at position $p$ gives $\{p\}\times[v_{\min}-\varepsilon,v_{\max}+\varepsilon]$; a Singleton $v$ is the case $v_{\min}=v_{\max}=v$, and a Pattern uses its value domain. Since positions can be uncertain, the windows of a construct $c$ are placed on its \emph{guaranteed extent}, the positions it covers under every feasible placement and length: $W_c$ holds the windows at the first and last guaranteed positions (one if they coincide), or, if no position is guaranteed, one window spanning all possible starts. Wildcards imply no window, but their lengths affect the positions of subsequent constructs in the query. In Example~\textcolor{TSseekMotivationProposal}{1} ($m=16$), Range~(a) gives a window at position~1, Pattern~(b) at~2 and~7, and Pattern~(c) at~8 and~12: readings 8--12 belong to the decline under every length, although it may extend to~14.

\textbf{Spatial Candidate Preservation.}
\textcolor{TSseekCurrentRevision}{Algorithm~\ref{alg:piecewise_approx} ensures that every reading lies within $\varepsilon$ of a segment covering its position, with no error accumulated across segments.} A reading in $[v_{\min},v_{\max}]$ at position $p$ makes that segment intersect $\{p\}\times[v_{\min}-\varepsilon,v_{\max}+\varepsilon]$. Every true match therefore meets every window of $W_c$. Different windows may be met by different segments. In Example~\textcolor{TSseekMotivationProposal}{1}, the probes at 2 and~7 can hit segments covering 1--6 and 7--12---so windows are combined on series IDs. Interior positions and the other constructs are checked during refinement; the segment filters of Sec.~\ref{sec:distExec} carry their own conservative bounds. Probe and segment-filter bounds are padded for floating-point rounding.

\begingroup\color{TSseekCurrentRevision}
\noindent\textbf{Construct Selection.} Let $\widehat{count}(w)$ be the sum of the segment counts of the grid cells touched by window~$w$ (Sec.~\ref{sec:index}). A segment can be counted in several cells, and a touched cell can contain segments that miss the window. Construct selection (Fig.~\ref{fig:Arch}, blocks~(c) and~(e)) uses
\[
\widehat{sel}(c)=\min_{w\in W_c}\widehat{count}(w),
\qquad c^*=\arg\min_c\widehat{sel}(c).
\]
Probing $c$ intersects the series-ID sets returned by the windows in $W_c$ (Alg.~\ref{alg:query_exec}, lines~\ref{ln:whole_window_ids}--\ref{ln:whole_intersect}). The intersection cannot be larger than the smallest set, so $\widehat{sel}(c)$, the minimum estimated count among those windows, gives a rough estimate of how many candidate series would remain.

\textsc{TSseek} verifies that the query uses the index's recorded grid and that the query's $\varepsilon$ is at least the index's approximation bound. If these checks pass, a zero count for a required window that fits within the grid's position and value bounds means no indexed segment meets that window, so the query has no matches. If the query proceeds to index lookup, only the windows in $W_{c^*}$ are executed (Sec.~\ref{sec:distExec} and Alg.~\ref{alg:query_exec}).
\par\endgroup

\par\endgroup

\subsection{Distributed Query Execution}
\label{sec:distExec}
\begingroup\color{TSseekMotivationProposal}
Alg.~\ref{alg:query_exec} summarizes query execution. After rewriting (line~\ref{ln:whole_parse}) and construct selection (Sec.~\ref{sec:selectivity}; line~\ref{ln:whole_select}), only the windows $W_{c^*}$ probe the index,
while the query constructs of $Q$ are enforced later by exact refinement against the complete original query.

\begingroup\color{TSseekMotivationProposal}
\begin{algorithm}[t]
 \color{TSseekMotivationProposal}\small
 \DontPrintSemicolon
 \KwIn{Dataset $\mathcal{T}$ with series of length $m$, INDEX$(\mathcal{T})$, query $Q$, grid statistics}
 \KwOut{Matching series IDs $F$}
 \Begin{
   $Q'\gets$ parse $Q$ and resolve feasible whole-series lengths and placements\nllabel{ln:whole_parse}\;
   \lIf{no feasible placement covers $m$ readings}{\Return{$\emptyset$}}
   \lIf{$Q'$ contains only Wildcards}{\Return{all series IDs in $\mathcal{T}$}}
   \ForEach{non-Wildcard construct $c$ in $Q'$}{
     $W_c\gets$ expanded windows at the ends of $c$'s guaranteed extent (one if equal), or one window spanning all possible starts\;
     $\widehat{sel}(c)\gets\min_{w\in W_c}\widehat{count}(w)$\;
     \lIf{grid parameters are verified and some $w\in W_c$ \textcolor{TSseekCurrentRevision}{fits within the grid's bounds and} has count zero}{\Return{$\emptyset$}}
   }
   $c^*\gets\arg\min_c\widehat{sel}(c)$\nllabel{ln:whole_select}\;
   $Cand\gets\emptyset$\;
   \ForEach{table $T$ among all $N_{\mathrm{tbl}}$ tables}{
     \ForEach{window $w\in W_{c^*}$}{
       $I_w\gets$ distinct series IDs whose segments intersect $w$ in $T$ and pass the applicable conservative filters\nllabel{ln:whole_window_ids}\;
     }
     $Cand_T\gets\bigcap_{w\in W_{c^*}} I_w$\nllabel{ln:whole_intersect}\;
     $Cand\gets Cand\cup Cand_T$\nllabel{ln:whole_union}\;
   }
   $F\gets\emptyset$\;
   \ForEach{raw series $t$ retrieved for $Cand$\nllabel{ln:whole_retrieve}}{
     \lIf{$\textsc{Refine}(Q,t)$}{add $t$'s ID to $F$\nllabel{ln:whole_refine}}
   }
   \Return{$F$}\;
 }
 \caption{\textsc{TSseek} Whole-Matching Query Processing Workflow.}
 \label{alg:query_exec}
\end{algorithm}
\par\endgroup

\textbf{Distributed Probing.}
All segments of a series are assigned to the same partition. The $N_{\mathrm{tbl}}$ partitions are stored in separate PostgreSQL tables across $M$ hosts (Sec.~\ref{sec:index}), with $N_{\mathrm{tbl}}>M$. Each partition is probed by an independent query \textcolor{TSseekRevisionRoundTwo}{(Fig.~\ref{fig:Arch}, block~(f))} in parallel, so multiple partitions per host permit concurrent table probes. 
\textcolor{TSseekCurrentRevision}{The driver sends the probe SQL for $W_{c^*}$ to all partitions in batches of concurrent queries.
The batch size bounds the number of simultaneous probe connections.
The returned series IDs stay in memory and pass directly to refinement within the same Spark application.}

\textbf{Conservative Segment Filtering.}
\textcolor{TSseekCurrentRevision}{A probe returns the IDs of series with at least one segment that intersects the window and passes the applicable segment filters.
For a Pattern with a direction or magnitude constraint, segment slopes can prune further using bounds that account for approximation error.
Requiring the slope to have the Pattern's sign could drop true matches.}
Suppose a segment with slope $s$ covers $r\geq1$ consecutive internal steps of the Pattern within its guaranteed extent. \textcolor{TSseekCurrentRevision}{Each such step satisfies $a\leq x_{j+1}-x_j\leq b$, where $a$ and $b$ are the lower and upper step bounds implied by the query.} Both readings at the ends of these steps lie within $\varepsilon$ of the segment (Sec.~\ref{sec:selectivity}), so
\[
a-\frac{2\varepsilon}{r}\leq s\leq b+\frac{2\varepsilon}{r}.
\]
\textcolor{TSseekCurrentRevision}{Without a magnitude, an increase gives $a=0$, i.e., $s\geq-2\varepsilon/r$, and a decrease gives $b=0$.
A flat Pattern without a magnitude gives step bounds $[-\tau,\tau]$ with absolute tolerance $\tau$.
With a magnitude constraint, the bounds follow from the Pattern's magnitude and direction.
The query tolerance can widen these bounds (Sec.~\ref{sec:Prelim}).
If the query implies only one step bound, the probe checks only the corresponding slope bound.
If the segment covers no full step within the guaranteed extent ($r=0$), the probe skips the slope test for that segment.}
Within each partition, the windows' ID sets are then intersected, and the per-partition results are unioned (lines~\ref{ln:whole_window_ids}--\ref{ln:whole_union}).

\textbf{Retrieval and Refinement.}
Raw time series objects \textcolor{TSseekRevisionRoundTwo}{(Fig.~\ref{fig:Arch}, block~(d))} are partitioned like the segments, i.e.,  each segment partition has a corresponding raw-series table holding the same series, with a B-tree on the series ID\eltabakhcomment{}. 
Candidate chunks are formed according to the configured parallelism and scheduled as Spark tasks.
\textcolor{TSseekCurrentRevision}{Each task fetches its candidates' raw series from the corresponding raw-series table using one SQL query per batch of IDs.
It checks each retrieved series against the complete original query and keeps only the matching IDs (Fig.~\ref{fig:Arch}, block~(g), and Alg.~\ref{alg:query_exec}, lines~\ref{ln:whole_retrieve}--\ref{ln:whole_refine}).}
Refinement is a dynamic program over (construct, start position) pairs: it matches constructs in order on the raw readings, trying each allowed length, and memoizes pairs that cannot complete a match to ensure each pair is expanded at most once~\cite{davis2021memoization}.
\par\endgroup

\subsection{Data-Driven \textcolor{TSseekCurrentRevision}{Parameter} Configuration}
\label{sec:DataDriven}
\begingroup\color{TSseekMotivationProposal}
\textsc{TSseek} sets two dataset-dependent parameters before preprocessing: segmentation threshold $\varepsilon$ and  grid geometry.

\textbf{Threshold $\varepsilon$.} \textcolor{TSseekCurrentRevision}{On a sample of series (e.g., 5\%),} we compute $\mathrm{span}_{90}=\mathrm{p}_{95}-\mathrm{p}_{5}$, the width of the central 90\% of the value distribution, which is robust to outliers and carries the data's units. We then set $\varepsilon=\alpha\cdot\mathrm{span}_{90}$ with $\alpha>0$. Since $\varepsilon$ bounds every segment's vertical error (Alg.~\ref{alg:piecewise_approx}), a smaller $\alpha$ typically yields more, finer segments. We calibrate $\alpha$ by running Alg.~\ref{alg:piecewise_approx}
\textcolor{TSseekCurrentRevision}{on the same sample} to target a mean of 7--9 segments per series, motivated by our ablation (Sec.~\ref{sec:exp}); the resulting $\alpha$ depends on the \textcolor{TSseekCurrentRevision}{dataset} and the record length.

\textbf{Grid Geometry.} The grid spans all reading positions and the data's global value bounds widened by $\varepsilon$, which contain every segment. \textcolor{TSseekCurrentRevision}{At the chosen $\varepsilon$, we initialize cell width and height from the sample's median horizontal and vertical segment spans. We adjust them to target median coverage of about two cells per axis, balancing localized counts against the cost of counting each segment in more cells.} The counts themselves are collected over all indexed segments during index construction (Sec.~\ref{sec:index}) and drive construct selection (Sec.~\ref{sec:selectivity}).
\par\endgroup

\makeatletter
\let\TSseekSavedStartSection\@startsection
\def\@startsection#1#2#3#4#5#6{%
  \let\@startsection\TSseekSavedStartSection
  \@startsection{#1}{#2}{#3}%
    {1.5ex minus 0.5ex}{0.7ex}{#6}}
\makeatother
\subsection{Subsequence-Matching Specialization}
\label{sec:SubseqProcessing}
\begingroup\color{TSseekMotivationProposal}
A subsequence query consists of $k\geq1$ sub-patterns, each matching a contiguous window of a series; a sub-pattern's start may be free, bounded, or fixed and its length fixed or variable, and for $k\geq2$, ordering and gap constraints relate the windows (Table~\ref{tab:subseq-classes}). Subsequence matching reuses the index, the partitioning, and raw-series retrieval of Secs.~\ref{sec:selectivity}--\ref{sec:distExec}. Index lookup and refinement are specialized as described below.

\textbf{Index Lookup.} Each sub-pattern $p_i$ yields necessary segment conditions, justified as in Sec.~\ref{sec:selectivity}: every reading of a match lies within $\varepsilon$ of its covering segment. Every sub-pattern admits the value-band condition $\phi_i$: a segment whose value range meets the sub-pattern's $\varepsilon$-expanded domain, plus the applicable conservative filters of Sec.~\ref{sec:distExec}. A fixed start also permits windows at the first and last guaranteed positions, as in whole matching.
\textcolor{TSseekCurrentRevision}{A bounded start permits one window over its possible starts.
In either single-pattern case, \textsc{TSseek} uses the windows or $\phi_i$, whichever its cost estimate favors.} For the multi-pattern classes (Q7--Q11), whose sub-pattern positions are relative, one statement per partition returns the series satisfying every $\phi_i$, possibly through segments at different positions: \texttt{GROUP BY} series ID \texttt{HAVING} $\texttt{bool\_or}(\phi_1)\wedge\dots\wedge\texttt{bool\_or}(\phi_k)$. Ordering, gaps, and the exact predicates are left to refinement. Under the grid-verification conditions of Sec.~\ref{sec:selectivity}, a zero count for a required window \textcolor{TSseekCurrentRevision}{that fits within the grid's bounds} proves an empty answer.

\eltabakhcomment{}
\textbf{\textcolor{TSseekRevisionRoundTwo}{Retrieval and Refinement.}} \textcolor{TSseekRevisionRoundTwo}{Raw-series retrieval follows Sec.~\ref{sec:distExec}.} For each candidate, the \textcolor{TSseekRevisionRoundTwo}{subsequence} refiner tests placement windows $[a,a{+}L{-}1]$ of each sub-pattern over its permitted starts and lengths: all $L$ readings must lie in its value domain, and its $L{-}1$ internal steps must satisfy its direction and any step constraint, with the tolerance of \textcolor{TSseekCurrentRevision}{Sec.~\ref{sec:Prelim}}. For a strictly increasing or decreasing sub-pattern with no step constraint and zero absolute \texttt{Threshold}, two prefix-sum arrays built in $O(m)$ time per series of length $m$ make each window test $O(1)$; other sub-patterns, such as the magnitude-constrained rise of Example~\textcolor{TSseekMotivationProposal}{1}, take $O(L)$ per window test, stopping at the first violation. When only series IDs are requested, a single-pattern search stops at its first valid window; a multi-pattern search first finds each sub-pattern's matching windows, then stops at the first combination that satisfies the ordering and gap constraints, which can be combinatorial in the worst case.
\par\endgroup

\section{Experimental Evaluation}
\label{sec:exp}

\subsection{Experimental Methodology and Setup}

\textbf{Cluster Setup \& Implementation.} Experiments run on a 112-core cluster (Intel Xeon E5-2690 v4, 512\,GB RAM/node, 7.3\,TB disk/node) with Debian, Spark 3.5.0, and PostgreSQL 15 with PostGIS for spatial indexing.

\textbf{Datasets.}
We evaluate \textsc{TSseek} using \textcolor{TSseekMotivationProposal}{four} datasets:

\textbf{(1) ECG (Electrocardiogram) Real-World Dataset.}
This dataset is derived from a large-scale 12-Lead Electrocardiogram Database for Arrhythmia Study (v1.0.0)~\cite{physionet_ecg}, which contains over 40,000 ECG recordings sampled 
at 500\,Hz across 12 leads, with detailed arrhythmia annotations.
From each recording, we extract leads I--III and aVR, convert 
amplitude measurements from µV to mV, and apply a 128-sample (0.256\,s) sliding window 
with about 75\% overlap to produce time-series segments.

\textbf{(2) Random Walk Benchmark Dataset.}
This is a widely adopted benchmark for evaluating time series analysis techniques~\cite{ShiehKeogh2009, Camerra2010iSAX2, Rakthanmanon2012, DPiSAX, DataSeriesIndexingGoneParallel, zhang2019tardis}.

 \textbf{(3) TSBS Fuel Consumption Synthetic Dataset.}
The Time Series Benchmark Suite (TSBS)~\cite{tsbs} provides synthetic telemetry data. 
We utilize the fuel consumption metric of simulated trucks to assess query scalability under industrial-like workloads.

\textcolor{TSseekMotivationProposal}{\textbf{(4) REFIT Electrical Load Real-World Dataset.}
REFIT~\cite{murray2017refit} contains real household active-power readings (in watts), sampled about every 8 seconds in 20 UK homes over two years.
We split each home's gap-free whole-house aggregate readings into non-overlapping 128-reading windows, keep the raw values, and replicate the windows with new identifiers to reach each target size.}

\textbf{Default Parameter Settings.}
All series have 128 readings, except in the length study (Fig.~\ref{fig:length-sensitivity}).
\textcolor{TSseekRevisionRoundTwo}{Following the calibration in Sec.~\ref{sec:DataDriven}, we set}
$\varepsilon_{\text{ECG}}=0.095$\,mV,
$\varepsilon_{\text{RW}}=1.59$,
$\varepsilon_{\text{TSBS}}=27.18$\textcolor{TSseekMotivationProposal}{, and
$\varepsilon_{\text{REFIT}}=65$}.
\eltabakhcomment{}
In the query processing experiments, each reported point is the average of five runs. 
Finally, as the dataset size increases, the number of partitions storing the data should also increase (as in distributed file systems by default). For our database setting, we use 40 tables for 25M series and double the number of tables as the  dataset size doubles.

\subsection{Baselines}
\label{sec:baselines}
We compare \textsc{TSseek} against four representative techniques:

\textbf{(1) Full Scan.} Spark applies \textsc{TSseek}'s exact matcher (Sec.~\ref{sec:distExec}) to every series without index lookup. The matcher tries permitted lengths for each construct in query order and memoizes failed (construct, start position) pairs. The matching series IDs serve as the reference answers for whole matching.

\textbf{(2) TARDIS.} \textsc{TARDIS}~\cite{zhang2019tardis} is a distributed iSAX-based index for whole-series similarity search on Spark. We adapt it to pattern queries by issuing $k$-NN searches with upper- and lower-bound numerical templates and a large $K$. We then check the retrieved candidates against the original pattern query.

\textbf{(3) SEAnet}~\cite{wang2021seanet}: a learned-embedding approach that trains an autoencoder; embeddings are converted to SAX symbols~\cite{lin2007experiencing} and indexed with an iSAX-style tree~\cite{ShiehKeogh2009, Camerra2010iSAX2} for $k$-NN search. Like \textsc{TARDIS}, \textsc{SEAnet} natively accepts only whole-series numerical queries. We use the same bounding-query adaptation for whole-matching pattern queries.

\textbf{(4) T-ReX}~\cite{T_Rex}: a segment-based subsequence engine that evaluates pattern queries directly over raw series without a persistent index. We use our Spark reimplementation of T-ReX for subsequence matching. We translate each query pattern into its T-ReX equivalent when possible.

\FloatBarrier

\subsection{Evaluation of Preprocessing and Index Construction}

\begin{figure}[t]
  \centering
  \includegraphics[width=\columnwidth]{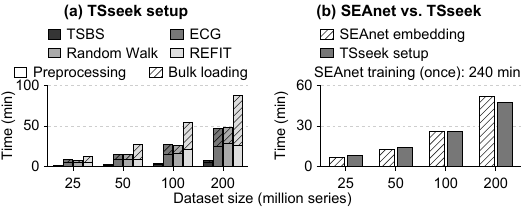}
  
  \caption{Setup time by dataset size. (a) \textsc{TSseek} preprocessing and bulk loading. (b) \textsc{SEAnet} embedding and \textsc{TSseek} setup.}
  \label{fig:tsseek-setup-3panel}
  
\end{figure}

\textbf{\textsc{TSseek} Overheads (Fig.~\ref{fig:tsseek-setup-3panel}(a)).}
Setup consists of Spark preprocessing and bulk loading into indexed PostgreSQL tables.
Preprocessing accounts for about 60\% of setup time on ECG and Random Walk and a larger share on TSBS.
On REFIT, bulk loading accounts for about 55--70\% of setup time, and total setup time grows approximately linearly with dataset size.

\textbf{TARDIS Comparison.} As presented next, since TARDIS is not designed for pattern-based queries, its recall drops to around 30\% (vs.\ \textsc{TSseek}'s exact answers) and its query response time is approximately an order of magnitude higher.

\textbf{SEAnet Comparison (Fig.~\ref{fig:tsseek-setup-3panel}(b)).}
\textsc{SEAnet}'s per-corpus embedding time is comparable to \textsc{TSseek}'s total setup time.
\textsc{SEAnet}'s one-time training takes 240 minutes and dominates its initial setup cost.

\begin{figure}[t]
  \centering
  \subfloat[By query type at 25M]{%
    \includegraphics[width=\columnwidth]{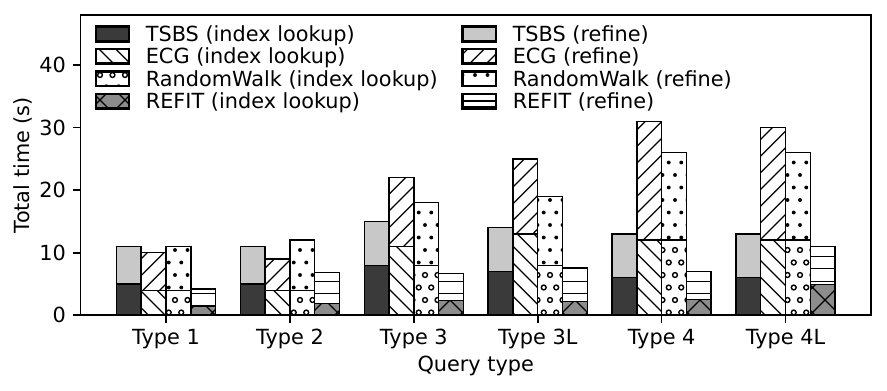}%
    \label{subfig:tsseek-total-by-type}
  }\\
  \subfloat[Per-type averages across four datasets, 25M--200M]{%
    \includegraphics[width=\columnwidth]{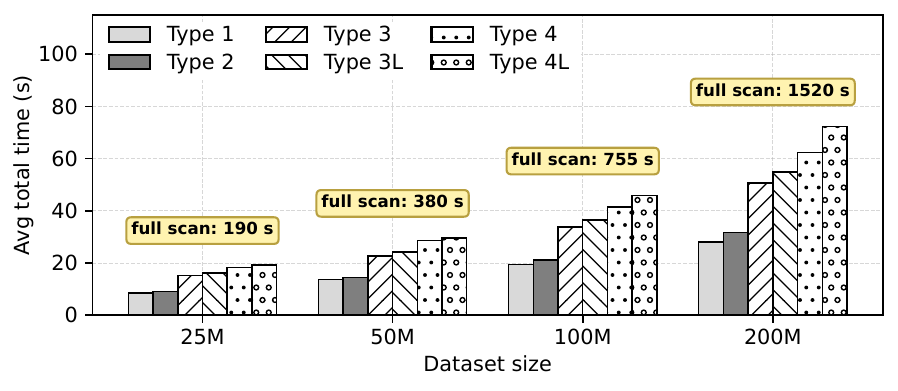}%
    \label{subfig:tsseek-avg-stacked}
  }
  
  \caption{\textsc{TSseek} whole-matching query time. Stacked bars show index lookup below retrieval and refinement (labeled ``refine''). Yellow labels show the full-scan baseline times.}
  \label{fig:tsseek-query-3panel}
\end{figure}

\subsection{Evaluation of Query Processing}

\textbf{Query Types (Whole-Matching)}.
The whole-matching workload uses four block categories:
\(\mathbf{S}\) (singleton value), \(\mathbf{R}\) (range for a single reading),
\(\mathbf{P_F}\) (fixed-length pattern block), and \(\mathbf{P_V}\) (variable-length pattern block).
In Types 1–4, pattern blocks are relatively \emph{short}; in the “L” variants (3L, 4L) pattern blocks are relatively \emph{long}.
We allocate the series length \(m\) among the four block categories in the stated proportions and instantiate the blocks accordingly.

{\small{}
\begin{itemize}
  \item \textbf{Type 1}: Balanced singletons/ranges/patterns, no variable-length patterns. Length allocation across \((S,R,P_F,P_V)\): \(\{\,\tfrac{1}{3},\,\tfrac{1}{3},\,\tfrac{1}{3},\,0\,\}\).

  \item \textbf{Type 2}: All four building block types are included.
  Length allocation: \(\{\,\tfrac{1}{4},\,\tfrac{1}{4},\,\tfrac{1}{4},\,\tfrac{1}{4}\,\}\).

  \item \textbf{Type 3}: No singletons; ranges plus patterns of both fixed and variable length (short).
  Length allocation: \(\{\,0,\,\tfrac{1}{3},\,\tfrac{1}{3},\,\tfrac{1}{3}\,\}\).

  \item \textbf{Type 3L}: Same composition as Type 3, but with fewer, longer patterns.
  Length allocation: \(\{\,0,\,\tfrac{1}{3},\,\tfrac{1}{3},\,\tfrac{1}{3}\,\}\).

  \item \textbf{Type 4}: Patterns only (no singletons or ranges), half fixed and half variable length (short).
  Length allocation: \(\{\,0,\,0,\,\tfrac{1}{2},\,\tfrac{1}{2}\,\}\).

  \item \textbf{Type 4L}: Same composition as Type 4 but with fewer, longer patterns.
  Length allocation: \(\{\,0,\,0,\,\tfrac{1}{2},\,\tfrac{1}{2}\,\}\).
\end{itemize}
}

\textbf{Per-Query-Type Performance (Whole-Matching).}
Fig.~\ref{subfig:tsseek-total-by-type} reports \textsc{TSseek}'s per-query-type total time at 25M across the \textcolor{TSseekMotivationProposal}{four} datasets (TSBS, ECG, Random Walk\textcolor{TSseekMotivationProposal}{, REFIT}). Query Types~1 and 2 include many singletons and ranges, giving strong point-wise selectivity, short candidate lists, and short processing time. In contrast, Types~3 and 4 are pattern-dominated (no singletons in 3/3L; patterns only in 4/4L), loosening early pruning and shifting more work to the refinement phase. Times change little for 3L/4L, except for Type~4L on REFIT.

\textbf{Scalability vs.\ Full Scan.}
Fig.~\ref{fig:tsseek-query-3panel}(b) reports the average total query time across the \textcolor{TSseekMotivationProposal}{four} datasets as dataset size grows from 25M to 200M for all six query types. \textsc{TSseek}'s total time grows linearly with dataset size and stays under 100\,s even at 200M.
The \textbf{full-scan} baseline (yellow labels) takes 190\,s at 25M and 1520\,s at 200M.
\textsc{TSseek} achieves approximately $20\times$ speedup at scale.

\begin{figure}[t]
  \centering
  \captionsetup[subfloat]{font=footnotesize,justification=centering,singlelinecheck=false,skip=2pt}
  \subfloat[\textsc{TSseek} vs.\ \textsc{TARDIS}]{%
    \includegraphics[width=0.85\columnwidth]{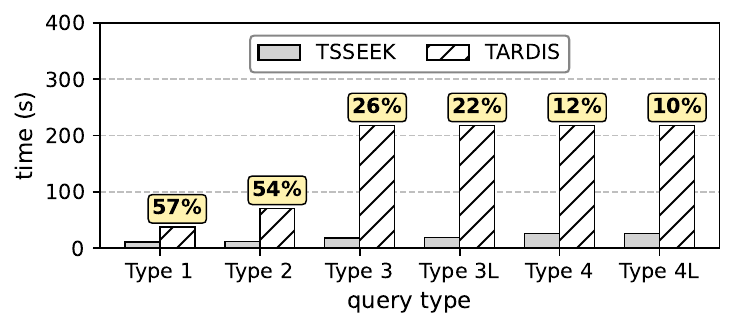}%
    \label{subfig:tardis-combined}
  }\\
  \subfloat[\textsc{TSseek} vs.\ \textsc{SEAnet}]{%
    \includegraphics[width=0.85\columnwidth]{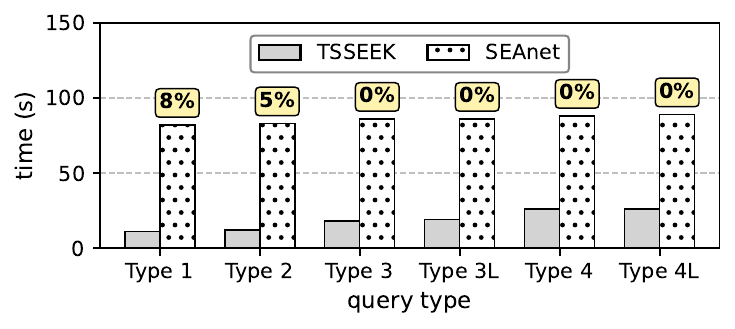}%
    \label{subfig:seanet-combined}
  }
  
  \caption{Whole-matching baseline comparison on Random Walk (25M) across six query types. Bars show query time; bubbles above each baseline bar show that baseline's recall (\%). \textbf{(a) \textsc{TSseek} vs.\ \textsc{TARDIS}.} \textbf{(b) \textsc{TSseek} vs.\ \textsc{SEAnet}.} \textsc{TSseek} recall = 100\% on all query types.}
  \label{fig:whole-baselines-combined}
\end{figure}

\textbf{Query Evaluation of \textsc{TSseek} vs.\ \textsc{TARDIS}.}
As described in Sec.~\ref{sec:baselines}, we adapt \textsc{TARDIS} via upper- and lower-bound queries on Random Walk (25M). We report recall as the fraction of reference matching series IDs returned after exact refinement. Fig.~\ref{fig:whole-baselines-combined}(a) shows the comparison.

\textsc{TARDIS} achieves moderate recall on point-wise Types~1--2 (54--57\%) but degrades sharply on pattern-dominated Types~3--4L (10--26\%): bounding queries cannot fully cover the search space defined by pattern predicates ($\mathbf{P_F}/\mathbf{P_V}$). Query time follows the same trend: tens of seconds on Types~1--2 but ${\sim}218$\,s on Types~3, 3L, 4, and 4L. After z-normalization, Types~1--2 segments (anchored by constants and narrow ranges) yield distinctive SAX words and small candidate sets, while pattern-dominated Types~3--4L segments collapse to similar SAX representations and enlarge the $k$-NN candidate set. \textsc{TSseek}, natively designed for pattern-based processing with efficient index retrieval and pruning, attains 100\% recall and stable query times (10.7--25.9\,s), achieving a $\mathbf{3.5\times}$--$\mathbf{12\times}$ speedup (largest gap on pattern-dominated types).

\textbf{Query Evaluation of \textsc{TSseek} vs.\ \textsc{SEAnet}.}
Following the same bounding-query adaptation with $K{=}100{,}000$, we evaluate \textsc{SEAnet} on the same six query types. \textsc{SEAnet}'s embeddings are queried via approximate nearest-neighbor search. Training \textsc{SEAnet} on 1M of 25M series required ${\sim}4$\,h, plus ${\sim}6.5$\,min to embed the full corpus.

Fig.~\ref{fig:whole-baselines-combined}(b) shows \textsc{SEAnet} achieves only 8\% and 5\% recall on Types~1 and 2, and 0\% on Types~3, 3L, 4, and 4L---learned embeddings cannot represent pattern-based queries. Query times are 82--89\,s across all types vs.\ \textsc{TSseek}'s 10.7--25.9\,s ($\mathbf{3.4\times}$--$\mathbf{7.7\times}$ speedup), with the gap narrowing as pattern content increases. \textsc{TSseek} keeps 100\% recall.

We now turn to the subsequence-matching workload, run over the same offline index via the \textcolor{TSseekMotivationProposal}{index-lookup and refinement} pipeline of Sec.~\ref{sec:SubseqProcessing}.

\textbf{Query Types (Subsequence-Matching).}
The workload has eleven classes (Table~\ref{tab:subseq-classes}). The single-pattern classes Q1--Q6 vary one Pattern's start constraint (free, bounded or fixed) and length (fixed or dynamic); the multi-pattern classes Q7--Q11 vary the order and gaps of multiple Patterns.

\textbf{Query Evaluation of Query Types (Subsequence-Matching).}
\begin{table}[t]
\centering
\footnotesize
\caption{Subsequence-matching query classes. $k$ defines the number of sub-patterns per query.}
\label{tab:subseq-classes}
\setlength{\tabcolsep}{4pt}
\begin{tabular}{@{}clll@{}}
\toprule
\textbf{Class} & \textbf{Group} & \textbf{Description} & \textbf{$k$} \\
\midrule
Q1  & single & FREE start, fixed length         & 1 \\
Q2  & single & FREE start, dynamic length       & 1 \\
Q3  & single & BOUNDED start, fixed length            & 1 \\
Q4  & single & BOUNDED start, dynamic length          & 1 \\
Q5  & single & FIXED start, fixed length              & 1 \\
Q6  & single & FIXED start, dynamic length            & 1 \\
Q7  & multi  & Adjacent (zero-gap, ordered)           & $\geq 2$ \\
Q8  & multi  & Fixed gap (ordered)                    & $\geq 2$ \\
Q9  & multi  & Range gap (ordered)                    & $\geq 2$ \\
Q10 & multi  & Flex gap, ordered                      & $\geq 2$ \\
Q11 & multi  & Flex gap, unordered                    & $\geq 2$ \\
\bottomrule
\end{tabular}
\end{table}

\begin{figure}[t]
  \centering
  \includegraphics[width=\columnwidth]{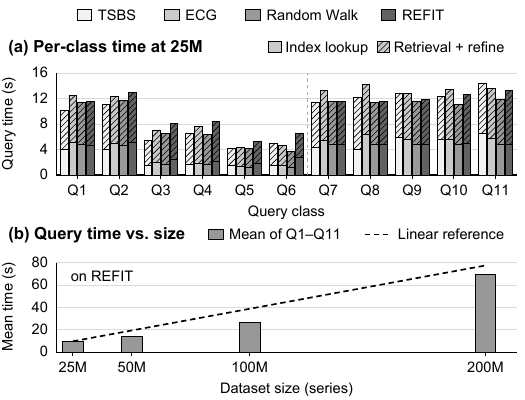}
  \caption{\textsc{TSseek} subsequence query time. (a) Per-class times at 25M on four datasets. (b) Mean time across all classes from separate scaling measurements on REFIT.}
  \label{fig:subseq-percase}
\end{figure}

Fig.~\ref{fig:subseq-percase}(a) shows \textsc{TSseek}'s per-class query times at 25M across four datasets. Fixed-start queries (Q5--Q6) are fastest on all four datasets, followed by bounded-start queries (Q3--Q4). Free-start and multi-pattern queries (Q1--Q2 and Q7--Q11) take about 10--14\,s. Index lookup retains fewer than 0.8\% of series for fixed-start queries and about 6--7\% for free-start and multi-pattern queries.

Fig.~\ref{fig:subseq-percase}(b) shows mean query time across all 11 classes as dataset size increases on REFIT. Mean time rises from 9.67\,s at 25M to 69.27\,s at 200M, a $7.2\times$ increase for an $8\times$ increase in dataset size. At each larger size, the measured time is below the dashed reference line, which shows proportional growth from the 25M measurement.

\begin{figure}[t]
  \centering
  \includegraphics[width=\columnwidth]{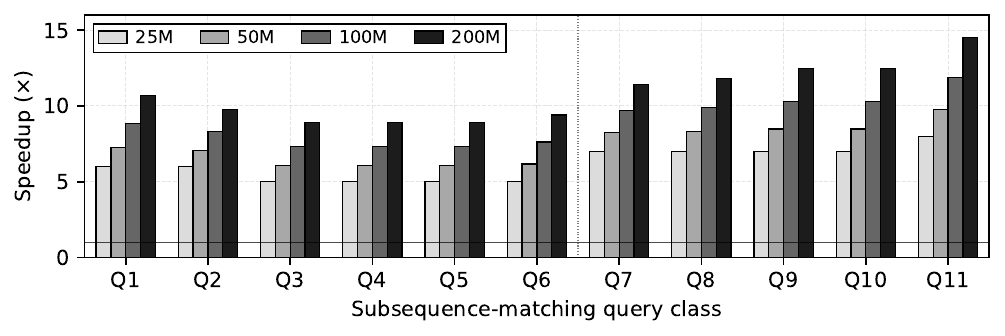}
  
  \caption{\textsc{TSseek} speedup over T-ReX on ECG as a function of query class and dataset size.}
  \label{fig:subseq-speedup}
  
\end{figure}

Fig.~\ref{fig:subseq-speedup} shows \textsc{TSseek}'s speedup over T-ReX~\cite{T_Rex} on ECG at four dataset sizes. Both systems use the same cluster configuration and data partitions. We validate returned matches by exact refinement. T-ReX permits zero changes in rising and falling Patterns. \textsc{TSseek} requires strict increases or decreases for those directions. At 25M, speedups range from $5\times$ to $8\times$ across all 11 classes. At 200M, they reach roughly 9--11$\times$ for single-pattern classes and 11--15$\times$ for most multi-pattern classes, with the largest gain for unordered flexible-gap matching (Q11). T-ReX scans the dataset and considers more window combinations when gaps are flexible. \textsc{TSseek}'s index prunes series before refinement.

\subsection{\textsc{TSseek} Ablation Studies}
\begin{figure}[!tb]
  \centering
  \includegraphics[width=\columnwidth,trim=0bp 6bp 0bp 6bp,clip]{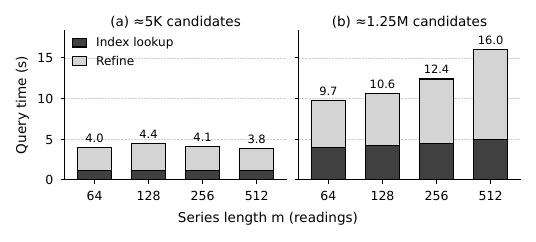}
  \caption{\textcolor{TSseekMotivationProposal}{Whole-matching query time versus series length on ECG-25M.}}
  \label{fig:length-sensitivity}
\end{figure}

We summarize three ablation studies.

\textbf{Selecting the Threshold ($\varepsilon$).}
The objective is to characterize how the segmentation threshold $\varepsilon$ (set via the scaling factor $\alpha$, see Sec.~\ref{sec:DataDriven}) trades off setup time against query time.
On TSBS-25M, sweeping $\alpha \in \{0.1, 0.25, 0.5, 0.6, 0.7, 0.75, 1.0\}$ reveals that total cost is minimized when the average segment count per series falls in $7$--$9$, achieved by $\alpha^{\ast} \approx 0.6$. For any new dataset, sampling a small slice and tuning $\alpha$ to hit this segment-count band yields a near-optimal operating point that balances preprocessing and query cost.

\textbf{Slope-Based Pruning.}
Our objective is to quantify the slope filter's impact on index lookup.
On TSBS-25M with a Type~4L query, we observe that
enabling slope filtering reduces the candidate set by about $123\times$ ($731{,}705 \to 5{,}926$) and cuts total query time from $24.7$\,s to $11.6$\,s, mostly by eliminating PostgreSQL serialization overhead on the larger candidate set.

\textbf{Sensitivity to Time-Series Length.}
\begingroup\color{TSseekMotivationProposal}
We vary $m \in \{64, 128, 256, 512\}$ on ECG, whose recordings are long enough to cut windows of every length from the same signals. We keep the dataset at 25M series and use queries with comparable candidate counts at each length. Fig.~\ref{subfig:tsseek-avg-stacked} covers dataset size. Calibration (Sec.~\ref{sec:DataDriven}) keeps the number of segments per series about the same at every length, so index size and setup barely change. Query time depends mainly on the \emph{candidates}, the series that pass index lookup and are checked on their $m$ raw readings.
At each length, we run 12 selective queries (about 5K candidates, typical) and 12 broad queries (about 1.25M candidates, 5\%), as shown in Fig.~\ref{fig:length-sensitivity}. Selective queries take about 4\,s at every length. For broad queries, total time rises from 9.7 to 16.0\,s. Refinement accounts for most of the increase, rising from 5.7 to 11.0\,s.
\par\endgroup

\section{Conclusion}
\label{sec:conclusion}
We proposed \textsc{TSseek}, a pattern-based search engine for distributed time series. Its query language uses regular-expression-inspired notation to specify trends, value ranges, and wildcards. \textsc{TSseek-X} indexes error-bounded line segments to prune candidate series before exact refinement on the original readings. Experiments on four real-world and benchmark datasets demonstrate exact pattern search at scales of up to 200 million time series.

\section*{AI-GENERATED CONTENT ACKNOWLEDGMENT}
AI-based tools (ChatGPT and Codex) were used for grammar correction, phrasing refinement, stylistic improvement, and editing plots based on author-provided measurements in Section VII. All conceptual content, analyses, experimental results, and technical contributions were fully conceived by the authors. No AI system was used to generate scientific ideas, references, data, or results.

{
\sloppy
\setlength{\emergencystretch}{3em} %
\bibliographystyle{IEEEtran}
\bibliography{tsseek_refs}
}

\end{document}